# Towards Atomic-Scale Control over Structural Modulations in Quasi-1D Chalcogenides for Colossal Optical Anisotropy

Guodong Ren[1*#], Shantanu Singh[2,3#], Gwan Yeong Jung[4#], Wooseon Choi[5], Huandong Chen[2], Boyang Zhao[2], Kevin Ye[2], Andrew R. Lupini[6], Miaofang Chi[6], Jordan A. Hachtel[6], Young-Min Kim[5,7], Jayakanth Ravichandran[2,3,8*], Rohan Mishra[1,4*]

[1] *Institute of Materials Science and Engineering, Washington University in St. Louis, St. Louis, MO 63130, USA*

[2] *Mork Family Department of Chemical Engineering and Materials Science, University of Southern California, Los Angeles, CA 90089, USA*

[3] *Core Center of Excellence in Nano Imaging, University of Southern California, Los Angeles, CA 90089, USA*

[4] *Department of Mechanical Engineering and Material Science, Washington University in St. Louis, St. Louis, MO 63130, USA*

[5] *Department of Energy Science, Sungkyunkwan University, Suwon 16419, Republic of Korea*

[6] *Center for Nanophase Materials Sciences, Oak Ridge National Laboratory, Oak Ridge, TN 37830, USA*

[7] *Center for 2D Quantum Heterostructures, Institute for Basic Science, Suwon 16419, Republic of Korea*

[8] *Ming Hsieh Department of Electrical Engineering, University of Southern California, Los Angeles, CA 90089, USA*

[#]These authors contributed equally: Guodong Ren, Shantanu Singh, Gwan Yeong Jung

*Corresponding authors: Guodong Ren (guodong.ren@wustl.edu); Jayakanth Ravichandran (j.ravichandran@usc.edu); Rohan Mishra (rmishra@wustl.edu)

**Keywords:** Quasi-1D chalcogenide, structural modulations, refractive index, charge ordering, optical anisotropy, electron microscopy

# ABSTRACT

Optically anisotropic materials are sought after for tailoring the polarization of light. Recently, colossal optical anisotropy ($\Delta n = 2.1$) was reported in a quasi-one-dimensional chalcogenide, $Sr_{9/8}TiS_3$. Compared to $SrTiS_3$, the excess Sr in $Sr_{9/8}TiS_3$ leads to periodic structural modulations and introduces additional electrons — that undergo charge ordering on select Ti atoms to form a highly polarizable cloud oriented along the *c*-axis, hence, resulting in the colossolal optical anisotropy. Here, further enhancement of the colossal optical anisotropy to $\Delta n = 2.5$ in $Sr_{8/7}TiS_3$ is reported through control over the periodicity of the atomic-scale modulations. The role of structural modulations in tuning the optical properties in a series of $Sr_xTiS_3$ compounds with $x = [1, 9/8, 8/7, 6/5, 5/4, 4/3, 3/2]$ has been investigated using density-functional-theory (DFT) calculations. The structural modulations arise from various stacking sequences of face-sharing $TiS_6$ octahedra and twist-distorted trigonal prisms, and are found to be thermodynamically stable for $1<x<1.5$. As $x$ increases, an indirect-to-direct band gap transition is predicted for $x \geq 8/7$ along with an increased occupancy of Ti-$d_{z^2}$ states. Together, these two factors result in a theoretically predicted maximum birefriengence of $\Delta n = 2.5$ for $Sr_{8/7}TiS_3$. Single crystals of $Sr_{8/7}TiS_3$ grown using a molten-salt flux method show $\Delta n \approx 2.5$, in excellent agreement with the predictions. Atomic-scale observations using scanning transmission electron microscopy confirm the feasibility of synthesizing $Sr_xTiS_3$ with varied modulation periodicities. While $Sr_{9/8}TiS_3$ crystals show a consistent modulation periodicity, $Sr_{8/7}TiS_3$ crystals show local pockets of $Sr_{6/5}TiS_3$, and underscore the need for further optimization of the synthesis conditions. Overall, these findings demonstrate compositonal tunability of optical properties in $Sr_xTiS_3$ compounds, and potentially in other hexagonal perovskites having structural modulations.

**Introduction**

Upon passing through a dielectric medium, the optical characteristics of light, including its speed, phase and intensity, are altered by the frequency-dependent complex index of refraction, $(n + i\kappa)$, of the medium, where $n$ is the refractive index and $\kappa$ is the extinction coefficient. Optically anisotropic materials have different index of refraction along different crystallographic orientations. Their anisotropy is characterized by birefringence ($\Delta n$) and dichroism ($\Delta \kappa$), which are, respectively, the difference between the refractive index and the extinction coefficient between two polarization directions[1, 2]. Optically anisotropic materials, particularly those having large and tunable $\Delta n$ and $\Delta \kappa$, are useful for generating and controlling polarized light, thereby enabling photonic applications through devices such as wave plates[3, 4] and optical sensors[5, 6].

Conventional birefringent crystals, such as calcite[1, 7] and rutile[8], show modest birefringence with $\Delta n < 0.3$ in spectral regions where they have high transparency. Recently, giant birefringence with $\Delta n > 0.5$, has been reported in several lower dimensional materials, including quasi-1D crystals and layered two-dimensional (2D) materials[9, 10]. Layered 2D materials have highly anisotropic bonding with weak interlayer van der Waals bonds and strong covalent intralayer bonds, which results in a large birefringence. h-BN and $MoS_2$ show $\Delta n \approx 0.7$ and 1.5 within their transparent region, respectively[9, 10]. However, harnessing the birefringence in layered 2D materials is challenging given that their optic axis is perpendicular to the weakly bonded layers. In these aspects, single crystals of quasi-1D perovskite chalcogenides are attractive. Niu *et al.* reported $\Delta n > 0.7$ in a broadband lossless spectral region spanning from mid- to long-wave infrared in single crystals of $BaTiS_3$.[11] These single crystals are mechanically robust compared to 2D van der Waals materials. They can also be grown along different orientations using scalable methods,[12, 13] thus providing different surface facets and optical axes with varying refractive index[13-18]. Recently, colossal optical anisotropy with $\Delta n \approx 2.1$ was reported in single crystals of another quasi-1D chalcogenide, $Sr_{9/8}TiS_3$[14]. Both $Sr_{9/8}TiS_3$ and $BaTiS_3$ have similar structures with chains of face-sharing $TiS_6$ polyhedra running along the long *c*-axis and arranged in a hexagonal pattern in the basal *ab*-plane. The Sr/Ba cations occupy the interchain interstitial positions. Compared to stoichiometric $BaTiS_3$, the excess Sr atoms in $Sr_{9/8}TiS_3$ modulate the $TiS_6$ chains into alternating blocks of octahedral and trigonal prismatic units with repeating periodicity. Furthermore, the additional electrons introduced by the excess $Sr^{2+}$ cations selectively localize on Ti atoms lying within the periodic modulations to form electron pockets that are oriented along the *c*-axis. These electron pockets selectively boost the polarizability along the *c*-axis, ultimately giving rise to the colossal birefringence[14]. Hexagonal perovskites, especially oxides, have also been reported to form modulated structures with varying stoichiometry and periodicity[19, 20]. Given the crucial role that the structural modulations play in enhancing the optical anisotropy in $Sr_{9/8}TiS_3$, a natural question that arises is whether the periodicity of the structural modulations and the excess electrons introduced by the Sr non-stoichiometry in $Sr_xTiS_3$ ($x \geq 1$) affect its optical anisotropy. If they do, then control over the modulation periodicity through synthesis methods[21] can enable access to crystals with varying birefringence.

In this Article, we report a record birefringence of $\Delta n = 2.5$ in $Sr_{8/7}TiS_3$ single crystals following computational predictions that show the tunability of optical anisotropy with control over atomic-scale structural modulations in $Sr_xTiS_3$. We used first-principles density-functional theory (DFT) calculations to investigate the thermodynamic stability and the optical properties of a series of modulated $Sr_xTiS_3$ structures. The calculations reveal that modulated structures with $x=[1, 9/8, 8/7, 6/5, 5/4, 4/3, 3/2]$ are thermodynamically stable, while $x = [1, 3/2]$ are above the convex hull. As $x$ increases from $SrTiS_3$, additional valence electrons from the $Sr^{2+}$ ions populate the Ti-$d_{z^2}$ states, which are nominally unoccupied in $SrTiS_3$. Two electronic factors are found to affect the optical anisotropy in this system: the direct or indirect nature of the bandgap and the occupancy of the Ti-$d_{z^2}$ states. The birefringence in $Sr_xTiS_3$ increases from $\Delta n = 2.1$ in the indirect bandgap compound $Sr_{9/8}TiS_3$ to a peak value of $\Delta n = 2.5$ in $Sr_{8/7}TiS_3$ — that has a direct bandgap —, followed by a monotonic decrease to $\Delta n = 1.2$ in $Sr_{4/3}TiS_3$. Subsequently, we report results of our experimental efforts to control the modulation periodicity in $Sr_xTiS_3$ by varying the stoichiometry. We employed two distinct synthesis methods to obtain $Sr_xTiS_3$ single crystals with two different stoichiometries. $Sr_{9/8}TiS_3$ single crystals were obtained using chemical vapor transport with iodine as a transport agent. $Sr_{8/7}TiS_3$ single crystals were grown using molten-salt flux method, with potassium iodide (KI) as the flux. Using aberration-corrected scanning transmission electron microscopy (STEM), we directly resolve the atomic structure and the modulation periodicity in $Sr_{9/8}TiS_3$ and $Sr_{8/7}TiS_3$ single crystals, respectively. Using Fourier-transform infrared spectroscopy (FTIR), we measured a record birefringence for $Sr_{8/7}TiS_3$, in excellent agreement with the DFT predictions. These findings highlight control over atomic-scale structural modulations in $Sr_xTiS_3$ as an effective strategy for tuning its optical anisotropy. Beyond establishing a new record birefringence in $Sr_{8/7}TiS_3$, our research underscores the untapped potential of modulated hexagonal perovksites — a vast and underexplored materials space — for achieving anisotropic physical properties.

## Results and Discussion

### Structural modulations in $Sr_xTiS_3$

With a chemical formula of $A_xMX'_3$ ($A$ = alkaline metal, $M$ = transition metal, $X'$ = anion), hexagonal perovskites form a series of structures having infinite chains of face-sharing $MX'_6$ polyhedra that are stacked in a hexagonal manner and are separated by the $A$-site cations. The $MX'_6$ polyhedral chains can modulate between octahedral and trigonal prismatic units with the periodicity of the modulations being either commensurate or incommensurate[19, 20, 22-24]. $Sr_xTiS_3$ compounds ($x > 1$) have been reported to be predominantly off-stoichiometric with $Sr_{9/8}TiS_3$ and $Sr_{8/7}TiS_3$ as the commonly observed modulations.[14, 25, 26]. These modulated $Sr_xTiS_3$ structures can be described as stacks of face-sharing $TiS_6$ octahedra (O), shown in blue in Figure 1a, and distorted trigonal prisms (T), shown in red. The arrangement and the ratio of these two building blocks are contingent upon the value of *x*.

Besides the experimentally reported modulated structures $Sr_{9/8}TiS_3$ and $Sr_{8/7}TiS_3$[14, 25, 26], we also constructed the following modulation periodicities based on their isostructural oxide counterparts: $Sr_{6/5}TiS_3$ ($Sr_{6/5}CoO_3$[27]), $Sr_{5/4}TiS_3$ ($Sr_{5/4}CoO_3$[28]), $Sr_{4/3}TiS_3$ ($Sr_{4/3}NiO_3$[29]), and $Sr_{3/2}TiS_3$ ($Sr_{3/2}CoO_3$[30]). We optimized all the structures using DFT and assigned space groups to the optimized structures using the ISOTROPY Software Suite[31]. Compared to stoichiometric $SrTiS_3$, the modulated $Sr_xTiS_3$ compounds have the Sr atoms displaced along the *ab*-plane and are accompanied by a twist distortion of $TiS_6$ units from octahedral to trigonal-prismatic configuration. The stacking sequence of the polyhedral building blocks (classified as O and T) defines the modulation periodicity of the $Sr_xTiS_3$ compounds. For instance, $Sr_{9/8}TiS_3$ has periodic $[-(T-O-T)-(O)_5-]_2$ repetition of 16 $TiS_6$ units within every 18 Sr layers along the *c*-axis, while $Sr_{6/5}TiS_3$ features 5 $TiS_6$ units with $[-(T)_2-(O)_3-]$ configuration repeating within every 6 Sr layers. Uniquely among the considered $Sr_xTiS_3$ compounds, $Sr_{8/7}TiS_3$ consists of two distinct stacking sequences involving $[-(T-O-T)-(O)_4-]$ and $[-(O)_5-(T)_2-]$ blocks, with each block containing 7 $TiS_6$ units.

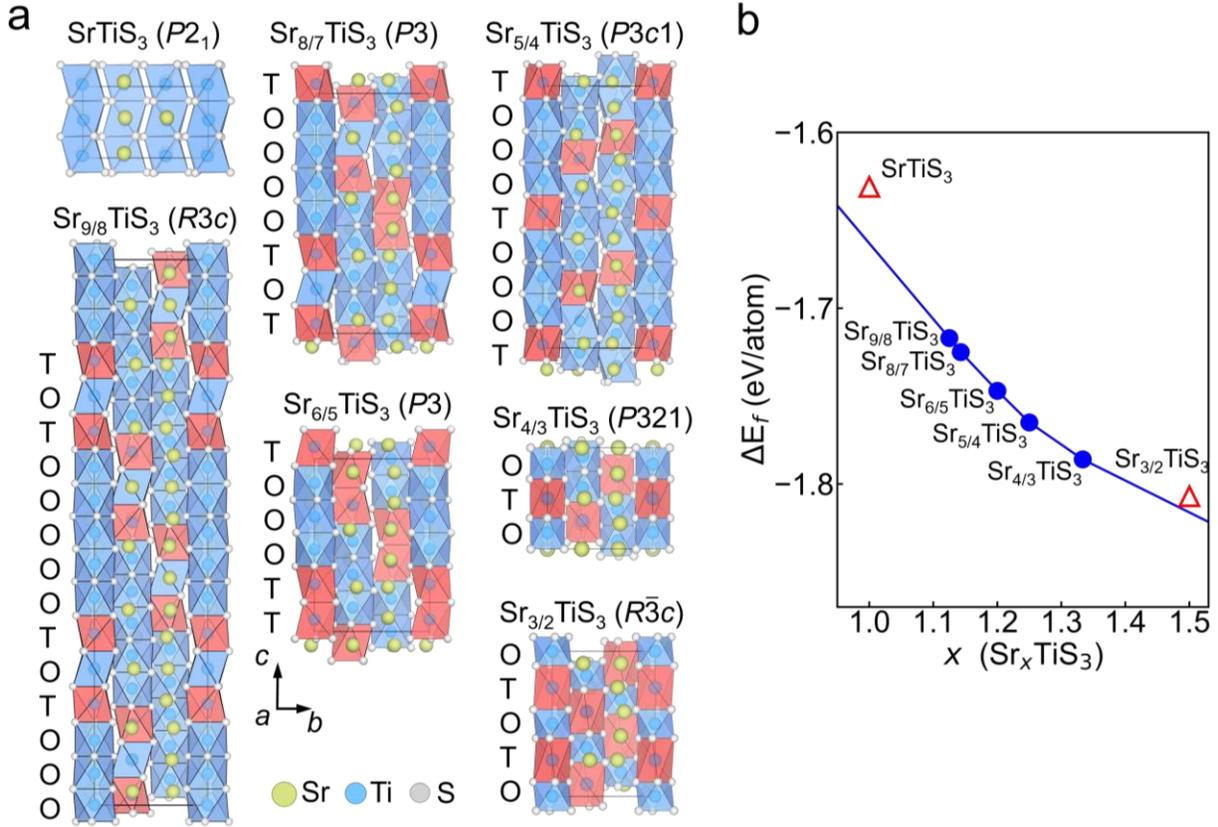

**Figure 1. Schematics of modulated $Sr_xTiS_3$ structures ($x$=[1, 9/8, 8/7, 6/5, 5/4, 4/3, 3/2]) and their thermodynamical stability**. **a,** Modulated structures of $Sr_xTiS_3$ compounds viewed along the [100] zone axis. The octahedral (O) and distorted trigonal prismatic (T) $TiS_6$ units are highlighted in blue and red, respectively. **b,** Thermodynamic convex hull diagram for $Sr_xTiS_3$ as a function of the Sr content, $x$. $SrTiS_3$ and $Sr_{3/2}TiS_3$ are metastable having formation energy above the hull.

## Thermodynamic stability of $Sr_xTiS_3$ compounds

To assess the thermodynamic stability of the $Sr_xTiS_3$ compounds, we constructed the convex hull with respect to the formation energy, $\Delta E_f$, of their possible decomposition products (see details in *Supporting Information Section II*). When a phase has a formation energy above the hull, it is considered metastable, as the system can lower its energy by decomposing into products that lie on the hull. We calculated the DFT total energy ($E_{total}$) of the $Sr_xTiS_3$ compounds and all the thermodynamically stable decomposition products, as listed in Table S1. The convex hull was then constructed through the grand canonical linear programming (GCLP) method[32] to identify the decomposition products that minimize the total energy. From the convex hull results shown in Figure 1b, modulated $Sr_xTiS_3$ structures with $1 < x < 1.5$ are thermodynamically stable, while $SrTiS_3$ ($x = 1$) and $Sr_{3/2}TiS_3$ ($x = 1.5$) are metastable having formation energy above the hull. Our thermodynamic stability analyses suggest that the modulation periodicity may be changed by growing $Sr_xTiS_3$ compounds with stoichiometry between $1 < x < 1.5$. One should also note that, due to the poor scaling of DFT calculations with system size, $Sr_xTiS_3$ structures having commensurate modulation periodicity (with finite unit cells) were only considered. The existence of incommensurate $Sr_xTiS_3$ structures with infinitely large unit cells, cannot be excluded. In fact, at high synthesis temperatures, the incommensurate structures are expected to be further stabilized by their higher configurational entropy.

**Optical anisotropy of $Sr_xTiS_3$ compounds**

To reveal the changes in optical properties with the structural modulations in $Sr_xTiS_3$ compounds discussed above, we calculated the frequency ($\omega$)-dependent dielectric function, $\varepsilon_{\parallel/\perp}(\omega) = \varepsilon_{1\parallel/\perp}(\omega) + i\varepsilon_{2\parallel/\perp}(\omega)$, for electric fields along the $c$-axis ($\parallel$) and perpendicular to it ($\perp$) using the formulation proposed by Gajdoš et al.[33] Using the computed dielectric function, we calculated the refractive index, $n(\omega) = \frac{\sqrt{2}}{2}\left[\sqrt{\varepsilon_1(\omega)^2 + \varepsilon_2(\omega)^2} + \varepsilon_1(\omega)\right]^{1/2}$, and the extinction coefficient, $k(\omega) = \frac{\sqrt{2}}{2}\left[\sqrt{\varepsilon_1(\omega)^2 + \varepsilon_2(\omega)^2} - \varepsilon_1(\omega)\right]^{1/2}$. The calculated optical properties of all the $Sr_xTiS_3$ compounds, excluding $Sr_{3/2}TiS_3$ which is computed to be metallic, are shown in Figure 2 and Figure S2 (*Supporting Information Section III*). From these plots, it is evident that the modulation periodicity in $Sr_xTiS_3$ has a significant effect on the dielectric response along the $c$-axis, as depicted by $n_\parallel(\omega)$ in Figure 2b, and $k_\parallel(\omega)$ in Figure 2e. In contrast, these modulations have a minimal impact on the *ab*-plane dielectric response, as shown in Figure 2a [$n_\perp(\omega)$], and Figure 2d [$k_\perp(\omega)$]. As the value of $x$ increases in $Sr_xTiS_3$, the refractive index along the $c$-axis, $n_\parallel(\omega)$, first increases from 4.83 ($x = 9/8$) to 5.40 ($x = 8/7$), then decreases monotonically to 3.70 ($x = 4/3$) within the lossless spectral region (6 to 16 μm). In contrast, the *ab*-plane refractive index, $n_\perp(\omega)$, remains relatively constant. This selective change of refractive index along the $c$-axis results in a change in the optical anisotropy with the modulation periodicity in these $Sr_xTiS_3$ compounds. Specifically, we find $\Delta n$ of $Sr_xTiS_3$ varies from 2.1 to 1.2 in the lossless spectral region with increasing $x$, as shown in Figure 2c. A similar trend is also observed for the linear dichroism ($\Delta \kappa$) of $Sr_xTiS_3$. With increasing Sr content, the extinction coefficient along the $c$-axis, $k_\parallel(\omega)$, shows a significant reduction, dropping from 3.97 at $x = 9/8$ to 1.38 at $x = 4/3$ within the spectral region of 1 to 4 μm. In contrast, the extinction coefficient along the *ab*-plane, $k_\perp(\omega)$, remains nearly unchanged around 0. This pronounced decrease in $k_\parallel(\omega)$, coupled with the constant value of $k_\perp(\omega)$, leads to a corresponding reduction in $\Delta \kappa$ from 3.86 to 1.37 with increasing $x$ in $Sr_xTiS_3$, as shown in Figure 2f.

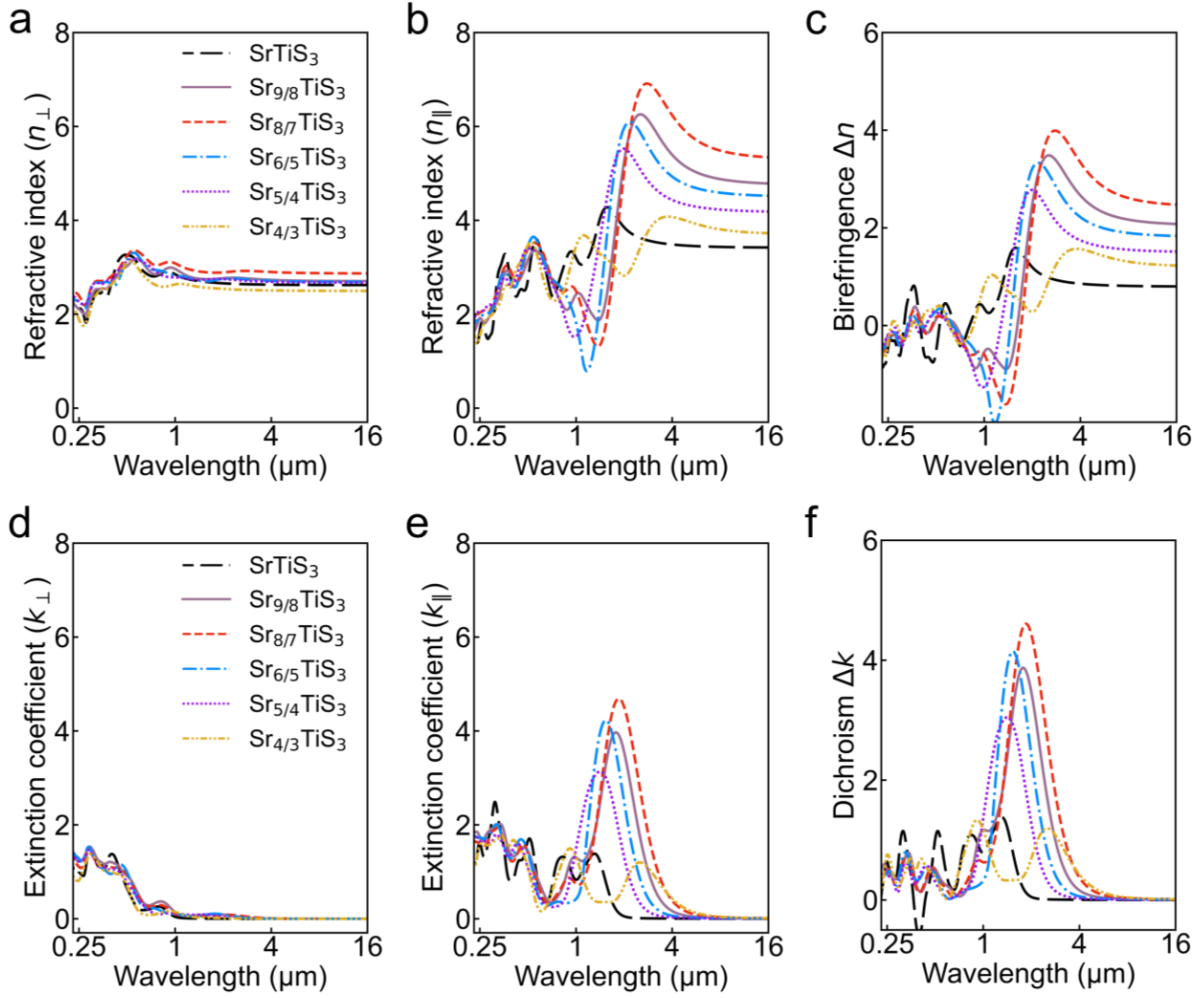

**Figure 2. DFT-calculated optical properties of modulated Sr$_x$TiS$_3$ compounds ($x$=[1, 9/8, 8/7, 6/5, 5/4, 4/3] ). a-b,** Calculated refractive indices of the Sr$_x$TiS$_3$ compounds for electric polarization along the *ab*-plane [$n_\perp(\omega)$] in **a** and along the *c*-axis [$n_\parallel(\omega)$] in **b**. **c**, Comparison of birefringence ($\Delta n$) of the Sr$_x$TiS$_3$ compounds considering the extraordinary direction parallel to the *c*-axis. **d-e**, Calculated extinction coefficients of the Sr$_x$TiS$_3$ compounds for electric polarization along the *ab*-plane [$k_\perp(\omega)$] in **d** and along the *c*-axis [$k_\parallel(\omega)$] in **e**. **f**, Comparison of dichroic spectra ($\Delta k$) of the Sr$_x$TiS$_3$ compounds.

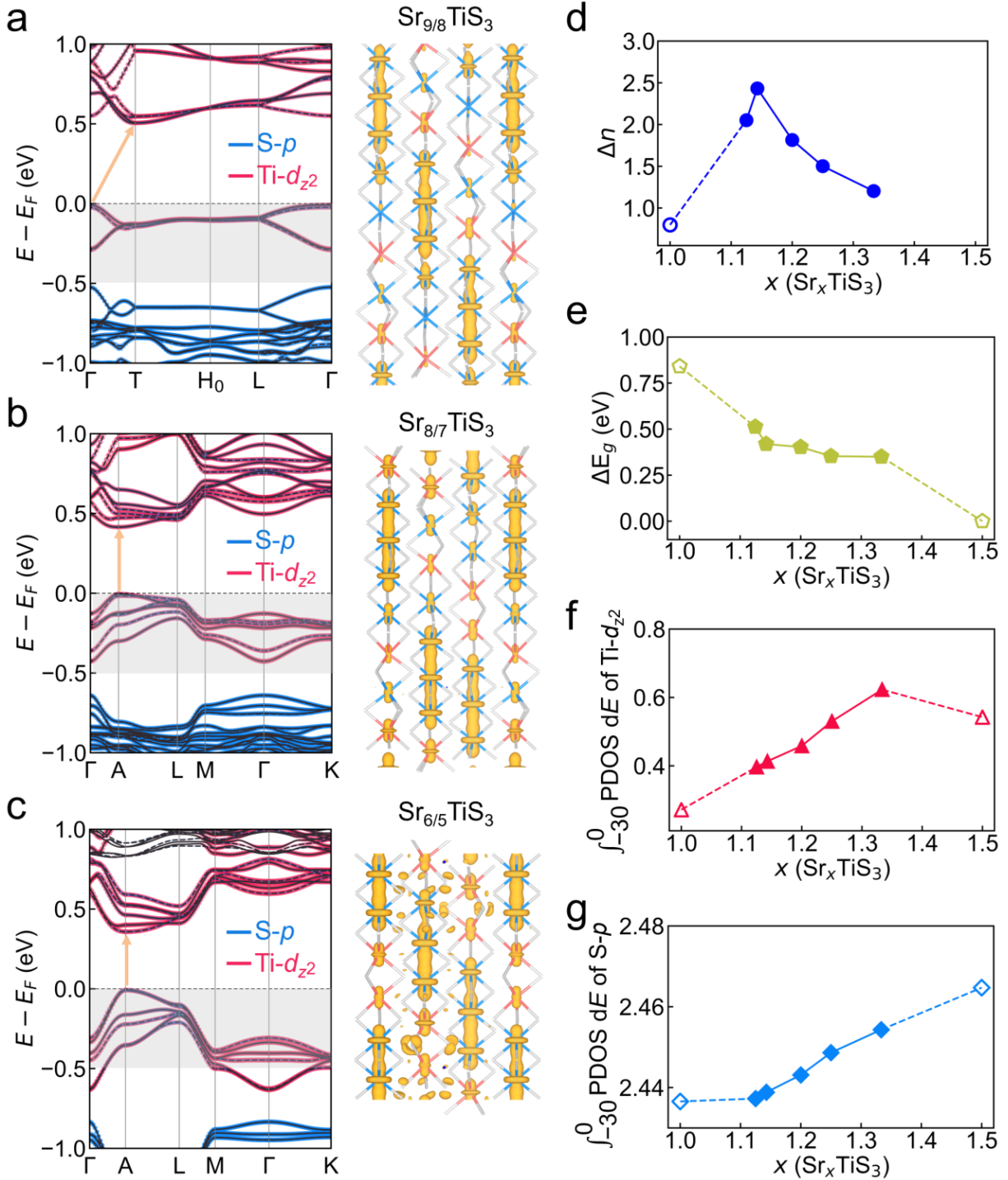

**Figure 3. Electronic structures of the $Sr_xTiS_3$ compounds ($x$=[9/8, 8/7, 6/5]) having different modulation periodicity. a-c,** Left panels: orbital-projected band structures for $Sr_{9/8}TiS_3$ in **a**, $Sr_{8/7}TiS_3$ in **b**, and $Sr_{6/5}TiS_3$ in **c**. The thicker red lines show the contribution from Ti-$d_{z^2}$ states, while the thicker blue lines correspond to S-$p$ states. The Fermi energy is set to 0 eV. The orange arrows, connecting the valence band maximum (VBM) to the conduction band minimum (CBM), indicate whether the bandgap is direct or indirect. Right panels: spatial distribution of the valence electrons within 0.5 eV below the Fermi energy

(highlighted with gray boxes in **a-c**). The isosurface is set to an electron density of 0.004 e/Å$^3$. **d,** Birefringence ($\Delta n$) of Sr$_x$TiS$_3$ as a function of $x$. **e,** Bandgap ($\Delta E_g$) of Sr$_x$TiS$_3$ as a function of $x$. **f-g,** Integrated orbital-projected density of Ti-$d_{z^2}$ states in **f** and S-$p$ orbital states in **g** as a function of $x$. The states were integrated from –30 eV to the Fermi energy at 0 eV. In **d-g**, metastable Sr$_{3/2}$TiS$_3$ and SrTiS$_3$ are depicted with hollow markers, while the stable compounds are denoted using filled markers.

**Electronic structure of Sr$_x$TiS$_3$ compounds**

The electronic structures of the Sr$_x$TiS$_3$ compounds were calculated and analyzed to understand the reason for the variation of $\Delta n$ with increasing $x$. Stoichiometric SrTiS$_3$ ($x = 0$) exhibits an indirect bandgap ($\Delta E_g = 0.84$ eV), characterized by a transition between the S-$p$ states in the valence band and the Ti-3$d$ states in the conduction band (as shown in Figure S3 *Supporting Information Section IV*). Stoichiometric SrTiS$_3$ is composed of face-sharing octahedral O units with $D3_d$ point-group symmetry and the Ti–S bonds pointing diagonally along the octahedral coordination. This coordination splits the Ti-3$d$ orbitals into two doublets ($e_g^\pi$ and $e_g^\sigma$) and a singlet ($a_{1g}$)[34]. The empty Ti-3$d$ states primarily contribute to the conduction bands, with the $a_{1g}$ state, having minimal overlap with the S-$p$ states, being the lowest in energy[14].

In the modulated Sr$_{9/8}$TiS$_3$ structure, the twist distortion of the TiS$_6$ octahedra is accompanied with changes in the Ti–Ti spacing along the $c$-axis from ~2.85 Å at O units (shown as blue in Figure 1a) to ~3.12 Å at the trigonal bipyramidal T units (shown as red in Figure 1a). These structural distortions transform the point group from $D_{3d}$ for the O units to ~$D_{3h}$ at the T units. Additionally, the variations in Ti coordination modify the splitting of the Ti-3$d$ states by lifting their degeneracies at different structural units. As a consequence, the Ti-$d_{z^2}$ states in original TiS$_6$ O units have been further split into selectively occupied $a_{1g}$ states at O units and empty $a_{1g}'$ states at T units (see the electronic structure and the isosurface plot of charge density in Figure 3a). The energy difference between the occupied $a_{1g}$ states and the empty $a_{1g}'$ states then reforms the bandgap in the modulated Sr$_{9/8}$TiS$_3$ structures. The character of the edge states in modulated Sr$_{9/8}$TiS$_3$ is in contrast with stoichiometric SrTiS$_3$ where the band edges are composed of occupied S-$p$ states and empty Ti-$d_{z^2}$ states (See Figure S3). In modulated Sr$_{9/8}$TiS$_3$, the selectively occupied Ti-$d_{z^2}$ states form a highly oriented electronic blob, exhibiting the strongest dipole transition to the empty Ti-$d_{z^2}$ states along the $c$-axis, which boosts the electric polarizability along the $c$-axis ($\varepsilon_{1\parallel}$).

We observe similar changes in the character of the band edges in other modulated Sr$_x$TiS$_3$ compounds, as can be seen in the orbital-projected band structure plots in Figure 3(a-c) (also see Figure S4 in *Supporting Information Section IV*). Additional electrons from excess Sr atoms preferentially occupy the Ti-$d_{z^2}$ states of the O units. The selective occupation of Ti-$d_{z^2}$ states can be visualized from the isosurface plots of the charge density as depicted in the right panels of Figure 3(a-c). As the ratio of Sr increases ($x$), more valence electrons from $Sr^{2+}$ ions occupy the Ti-$d_{z^2}$ states hybridized with S-$p$ states — that are mostly unoccupied in SrTiS$_3$. Increasing $x$ also transforms the indirect bandgap in SrTiS$_3$ and Sr$_{9/8}$TiS$_3$ to a direct bandgap in Sr$_x$TiS$_3$ with $x$=8/7, 6/5, and 5/4. Concurrently, the bandgap decreases monotonically from 0.84 eV in SrTiS$_3$ to 0.35 eV in Sr$_{4/3}$TiS$_3$, until Sr$_{3/2}$TiS$_3$ shows metallic character, as shown in Figure 3e. We calculated the occupation of the Ti-$d_{z^2}$ and S-$p$ states by integrating their projected density (see PDOS in Figure S5 in *Supporting Information Section IV*) below the Fermi energy. The results are shown in Figures 3(f-g). We find that with increasing $x$ in Sr$_x$TiS$_3$, the occupation of Ti-$d_{z^2}$ and S-$p$ states undergo a noticeable increase to accommodate the additional valence electrons from the excess Sr ions. Overall, the increasing occupation of Ti-$d_{z^2}$ and S-$p$ states is approximately inversely correlated with the decreasing birefringence $\Delta n$, as shown in Figure 3d, confirming the critical role of these states in modulating optical anisotropy in Sr$_x$TiS$_3$. In modulated Sr$_{9/8}$TiS$_3$, which has an indirect bandgap, the band-to-band transition strength, as determined by the dipole transition matrix elements[14], reveals that the dipole transitions around the band edge from the selectively occupied Ti-$d_{z^2}$ states to unoccupied Ti-$d_{z^2}$ states lead to a significant enhancement of the dielectric polarizability along the $c$-axis ($\varepsilon_{1\parallel}$) compared to SrTiS$_3$, where the transitions occur from occupied S-$p$ states to empty Ti-$d_{z^2}$ states (See Figure 2 and Figure S2). Given the similar band-edge characteristics of other modulated Sr$_x$TiS$_3$ compounds, it is anticipated that their dielectric function parallel to the $c$-axis ($\varepsilon_{1\parallel}$) are also predominantly determined by dipole transitions from the selectively occupied to unoccupied Ti-$d_{z^2}$ states. However, as more Sr ions are introduced in modulated Sr$_x$TiS$_3$, two concurrent effects occur. On the one hand, the bandgap evolves from indirect in SrTiS$_3$ and Sr$_{9/8}$TiS$_3$ to direct with $x$= 8/7, 6/5 and 5/4, which enhances the direct transitions from valence band to conduction band[35]. On the other hand, the increased Sr content leads to more occupied Ti-$d_{z^2}$ states, thereby reducing the number of available unoccupied Ti-$d_{z^2}$ states in the conduction band. Consequently, the allowed transitions involving Ti-$d_{z^2}$ states along the $c$-axis are suppressed with increasing $x$ in Sr$_x$TiS$_3$. These two critical factors—the nature of the bandgap and the occupancy of the Ti-$d_{z^2}$ states —govern the refractive index $n_\parallel(\omega)$ along the $c$-axis (Figure 2b) and the birefringence $\Delta n$. Specifically, $n_\parallel(\omega)$ and $\Delta n$ initially increase when going from Sr$_{9/8}$TiS$_3$ to Sr$_{8/7}$TiS$_3$, then decrease monotonically at higher $x$ in Sr$_x$TiS$_3$ (Figure 3d). Overall, Sr$_{8/7}$TiS$_3$ with direct bandgap shows the optimal optical anisotropy as compared to other Sr$_x$TiS$_3$ compounds.

**Direct observation of structural modulations with varying periodicity in Sr$_x$TiS$_3$ crystals**

Based on the theoretical calculations, it is evident that Sr$_{9/8}$TiS$_3$ and Sr$_{8/7}$TiS$_3$ compositions demonstrate the highest observed optical anisotropy. While Sr$_x$TiS$_3$ has been reported to crystallize in powder form through $1.05 \leq x \leq 1.22$ composition range,[36] single crystals have been reported with only two compositions ($Sr_{9/8}TiS_3$[25, 37] and $Sr_{8/7}TiS_3$[38]). Since it is evident from theoretical calculations that $Sr_{9/8}TiS_3$ and $Sr_{8/7}TiS_3$ compositions demonstrate the highest optical anisotropy, we have experimentally explored the potential for controlling the modulation periodicity of $Sr_xTiS_3$ by employing two distinct growth methods for the two reported stoichiometries. Single crystals of $Sr_{9/8}TiS_3$ were grown using the chemical vapor transport method, as reported elsewhere.[14] We used the molten flux method to obtain the single crystals of $Sr_{8/7}TiS_3$. First, we prepared the precursor powders by the $CS_2$ sulfurization of the commercially available $SrTiO_3$ powders. Then, single crystals of $Sr_{8/7}TiS_3$ were grown using KI as the flux[12] using a 1:40 powder-to-flux ratio. Long, needle-like crystals around a few mm in length and with thickness typically ranging from 25-100 $\mu$m were obtained. Further details on the underlying thermodynamic and kinetics factors influencing the stabilization of different stoichiometries as well as detailed analysis of single crystal XRD (SC-XRD) measurements used to confirm the bulk crystal structure will be discussed in a separate work.

To directly observe the structural modulations in the two $Sr_xTiS_3$ ($x = 9/8, 8/7$) compounds, we performed atomically resolved imaging using an aberration-corrected STEM. High-angle annular dark field (HAADF)-STEM images of the two $Sr_xTiS_3$ crystals viewed along the [100] zone axis are shown in Figures 4(a-d). In this imaging mode, the intensity of the atomic columns is approximately proportional to the square of the effective atomic number of the columns ($Z^2$)[39]. When projected along the [100] zone axis, the Ti and Sr columns overlap within −(T−O)− or −(T−T)− building blocks in different modulated $Sr_xTiS_3$ structures, as shown in Figure S1 (*Supporting Information Section I*). In HAADF-STEM images, these overlapped Ti and Sr columns exhibit higher intensity compared to the Sr-only atomic columns, enabling their distinction based on contrast[14]. This distinctive structural feature is pivotal in our STEM analysis for identifying the structural units and determining the corresponding modulation periodicities in $Sr_xTiS_3$ compounds.

The $Sr_{9/8}TiS_3$ single crystals display a well-defined modulation periodicity, as demonstrated by the large field-of-view HAADF image in Figure 4a and the corresponding fast Fourier transform (FFT) pattern in the inset. The superlattice spots labeled in the FFT pattern reveal the periodic structural modulation along the *c*-axis, recurring every 18 atomic layers. This observed modulation periodicity matches our calculated $Sr_{9/8}TiS_3$ structure, as shown in Figure 1a and Figure S1. In the high-magnification HAADF image shown in Figure 4b, we observe a periodic arrangement of brighter triplets, which arise from the overlap of Ti and Sr columns within the triple blocks of –(T–O–T)– along the [100] projection (see Figure S1). By counting the stacking sequence of building blocks (O and T units) from the intensity variations in a line profile, as shown in Figure 4c, we identified $TiS_6$ chains have a periodic sequence of $[-(T-O-T)-(O)_5-]_2$ along the *c*-axis. This stacking sequence defines a modulation periodicity of 16 units of $TiS_6$ within every 18 Sr layers, which further corroborates the modulation periodicity existing in $Sr_{9/8}TiS_3$.

The $Sr_{8/7}TiS_3$ crystals also display a long-range modulation periodicity in the HAADF image (Figure 4c), with the FFT pattern (inset) showing superlattice spots that indicate a structural modulation along the *c*-axis. In the high-magnification HAADF image shown in Figure 4d, we observe two distinct stacking sequences involving [−(T−O−T)−(O)$_4$−] and [−(O)$_5$−(T)$_2$−] blocks, with each block containing 7 $TiS_6$ units. This structural modulation agrees well with our calculated $Sr_{8/7}TiS_3$ structure (Figure 1a and Figure S1). Furthermore, a detailed comparison of the intensity and spacing between atomic columns in both the experimental and simulated HAADF images, as evidenced by the line profiles in Figure 4d, further validates the local modulation periodicities in $Sr_{8/7}TiS_3$. We also noted that some local regions of the synthesized $Sr_{8/7}TiS_3$ crystals exhibit aperiodic structural modulations, characterized by brighter horizontal columns (perpendicular to the *c*-axis) that segment the crystal into distinct domains, as shown in HAADF-STEM images (Figure S6-S7 in *Supporting Information Section V*). Structural template-matching analyses confirm a mixture of local modulation periodicities of $Sr_{8/7}TiS_3$ and $Sr_{6/5}TiS_3$ in these regions. This local aperiodicity could be a result of deviation from the stoichiometry, complicating the stabilization of the $Sr_{8/7}TiS_3$ phase and underscoring the need to optimize synthesis conditions to achieve purer structural modulations.

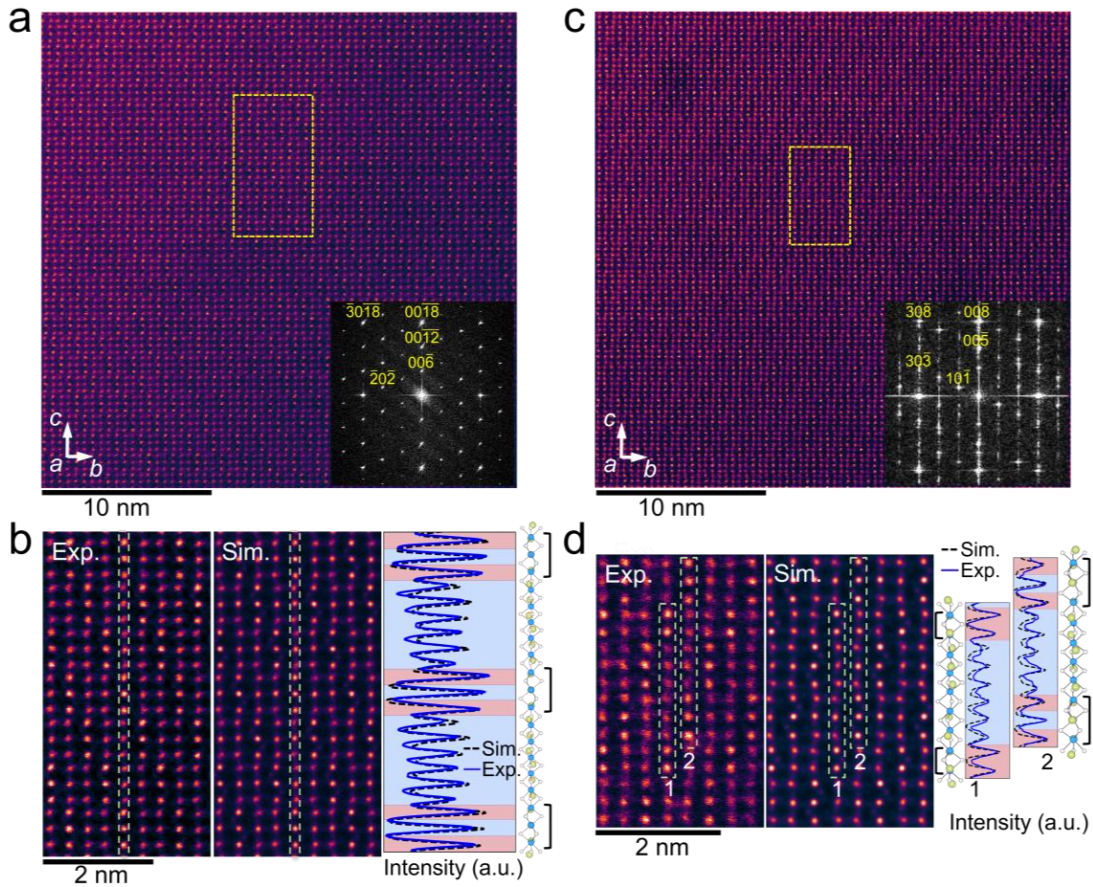

**Figure 4. Atomic-resolution HAADF-STEM images of Sr$_x$TiS$_3$ crystals with periodic and aperiodic modulations viewed along the [100] axis. a,** Large field-of-view HAADF-STEM image of a Sr$_{9/8}$TiS$_3$ crystal synthesized using chemical vapor transport. The inset shows the corresponding FFT pattern, which matches with the modulation periodicity of Sr$_{9/8}$TiS$_3$. **b,** High-magnification experimental (leftmost) and simulated (center) HAADF-STEM images of a Sr$_{9/8}$TiS$_3$ crystal showing the repeating brighter triplets at –(T–O–T)– blocks. In the rightmost panel, the line profile shows the intensity variation across the atomic columns within the yellow dashed box in the HAADF image, where red and blue shaded regions correspond to (T) units and (O) units, respectively. A schematic of one column of atoms within Sr$_{9/8}$TiS$_3$ is presented on the right where the –(T–O–T)– blocks are highlighted with square brackets. **c,** Large field-of-view HAADF-STEM image of a Sr$_{8/7}$TiS$_3$ crystal synthesized using molten-flux method. The inset shows the corresponding FFT pattern with superlattice spots indicating the structural modulation along the *c*-axis. **d,** High-magnification experimental (leftmost) and simulated (center) HAADF-STEM images of the yellow dashed box in (c), showing two distinct stacking sequences of brighter doublets and triplets along the *c*-axis: (1) [−(O)$_5$−(T)$_2$−] and (2) [−(T−O−T)−(O)$_4$−] stackings. The rightmost panel displays the intensity variation profiles across the two atomic columns marked by the dashed green boxes in the HAADF images. The (T) and (O) units are indicated with red and blue shaded backgrounds, respectively. Structural schematics of [−(O)$_5$−(T)$_2$−] and [−(T−O−T)−(O)$_4$−] stacking configurations within Sr$_{8/7}$TiS$_3$ are presented on the right.

## Optical anisotropy measurements

To investigate the optical properties of the $Sr_{8/7}TiS_3$ crystals, polarization-resolved FTIR measurements were performed at room temperature in the reflection mode. The polarization of the incident light was kept parallel and perpendicular to the crystal's uniaxial optic axis ($c$-axis), yielding anisotropic reflectance spectra (Figure 5a). The large optical anisotropy is evident from the significant difference observed in the reflectance for the two polarizations. The reflectance spectra collected for both orientations exhibit Fabry-Pérot interference fringes, originating from the interference between the light reflected from the top and the bottom surface of the measured single-crystal sample. The extinction of the Fabry-Pérot fringes at shorter wavelengths indicates the onset of the absorptive regime. Our DFT calculations demonstrate that the refractive index is relatively constant in the transparent regime. Thus, the periodicity of the interference fringes can be used to approximate the spectrum-averaged real part of the refractive index ($n$) for the extraordinary ($\parallel c$) and ordinary ($\perp c$) optical directions. The frequency, $\nu$, for such fringes, in case of normal incidence, is related to the refractive index of the material ($n$) and the thickness of the sample ($d$), by the relation $\nu = 2nd$.[40, 41] The FFT-derived refractive index values are ~5.1 and ~2.6 for the extraordinary and the ordinary directions, respectively, as shown in Figure 5b, giving a birefringence ($\Delta n$) value of ~2.5. To confirm the refractive index values, a two-term Sellmeier equation-based model[42] was used to approximate the wavelength dispersion for the refractive index. The fitting for the model was done to match the periodicity of the fringes observed in the reflectance spectra. The refractive index values from the model match well with the FFT-derived values (*Supporting Information, Section VII*). Figure 5c compares the calculated and experimental birefringence values for $Sr_xTiS_3$ crystals in the lossless spectral region, showing excellent agreement for both $Sr_{9/8}TiS_3$[14] and $Sr_{8/7}TiS_3$ modulated compounds. Both the theoretical calculations and experimental measurements demonstrate a larger optical anisotropy in $Sr_{8/7}TiS_3$ compared with other modulated $Sr_xTiS_3$ compounds.

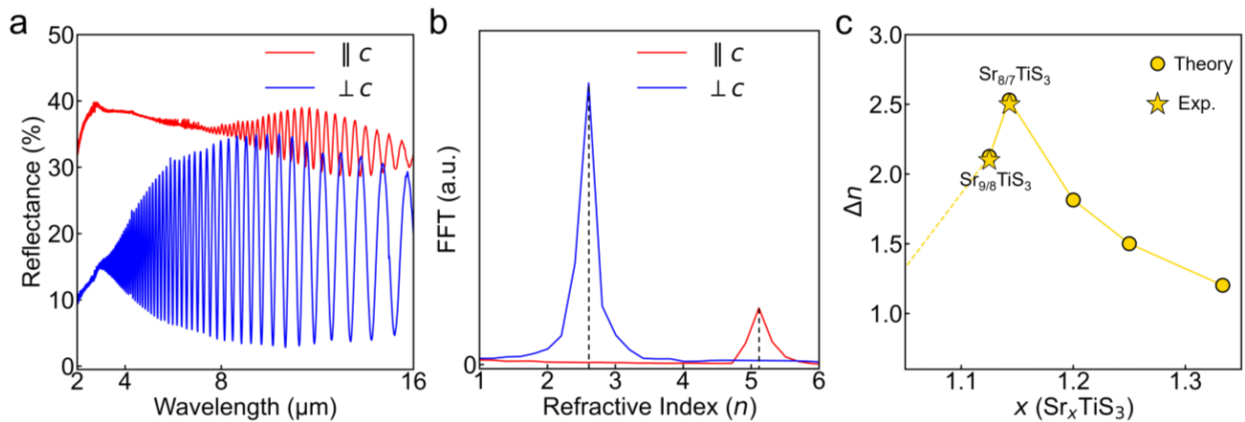

**Figure 5. Polarization-resolved FTIR measurements. a,** FTIR reflectance spectra for a $Sr_{8/7}TiS_3$ crystal showing large anisotropy between the ordinary (⊥c) and the extraordinary (∥c) polarization. **b,** FFT analysis on the reflectance spectra for the polarization-dependent refractive index of a $Sr_{8/7}TiS_3$ crystal. **c,** Comparison between the experimental and computed birefringence of modulated $Sr_xTiS_3$ compounds. The birefringence for $Sr_{9/8}TiS_3$ is taken from the literature.[14]

## Conclusion

In this work, we have demonstrated that the periodicity of atomic-scale structural modulations in $Sr_xTiS_3$ structures systematically changes their refractive index and optical anisotropy. DFT calculations suggest that modulated $Sr_xTiS_3$ structures with $1 < x < 1.5$ lie on the convex hull, and are expected to be stable against decomposition. With increasing $x$ in modulated $Sr_xTiS_3$, two concurrent effects occur: first, the bandgap evolves from indirect in $SrTiS_3$ and $Sr_{9/8}TiS_3$ to direct for $x=[8/7, 6/5, 5/4]$, which enhances the direct transitions from the valence band to the conduction band; second, as more Ti-$d_{z^2}$ states become occupied, they result in a reduction of dipole transitions between occupied and empty Ti-$d_{z^2}$ states across the band gap. Together, these two factors—the transformation of the bandgap and the increased occupancy of Ti-$d_{z^2}$ states—determine the refractive index $n_\parallel(\omega)$ along the $c$-axis, and the birefringence $\Delta n$. Specifically, $n_\parallel(\omega)$ and $\Delta n$ increase when moving from $Sr_{9/8}TiS_3$ to $Sr_{8/7}TiS_3$, then decrease monotonically for higher $x$ in $Sr_xTiS_3$. Overall, $Sr_{8/7}TiS_3$ having direct bandgap exhibits the most favorable optical anisotropy ($\Delta n \approx 2.5$) among the $Sr_xTiS_3$ compounds within the lossless spectral range of 6 to 16 μm. We were able to synthesize $Sr_{9/8}TiS_3$ and $Sr_{8/7}TiS_3$ crystals using two distinct synthesis methods. Direct observation of the atomic structure using STEM imaging reveals a consistent modulation periodicity in $Sr_{9/8}TiS_3$, whereas the $Sr_{8/7}TiS_3$ crystals have local regions with a mixture of $Sr_{8/7}TiS_3$ and $Sr_{6/5}TiS_3$ periodicities, highlighting the need to optimize the synthesis conditions for purer structural modulations. By correlating atomic-scale modulations with macroscopic anisotropy, we provide a framework for designing material candidates with variable properties for targetted photonic applications such as optical communication and polarized imaging. The demonstration of record $\Delta n$ in $Sr_{8/7}TiS_3$ underscores the broader potential of modulation engineering in other hexagonal perovskites,[43] where nanoscale structural complexity can be harnessed to achieve anisotropic optical responses that are inaccessible in conventional crystals. Beyond photonic applications, such modulations may also result in changes in other anisotropic physical properties, involving phonons, and electrons.[13]

## Methods

### Crystal growth

Single crystals of $Sr_{9/8}TiS_3$ were grown via the chemical vapor transport method using iodine as the transport agent. Strontium sulfide powder (Alfa Aesar, 99.9%), titanium powder (Alfa Aesar, 99.9%), sulfur pieces (Alfa Aesar, 99.999%), and iodine pieces (Alfa Aesar 99.99%) were stored and handled in a nitrogen-filled glove box. SrS, Ti, and S in a 1: 1: 2 ratio, with a total weight of 1.0 g, were added to a quartz ampoule along with ~0.75 mgcm$^{-3}$ of iodine. The quartz ampoule was evacuated and sealed using a blow torch. The sealed ampoule was then loaded into a two-zone furnace, where it was heated to the reaction temperature of 1055 °C at 100 °C/h and held for 150 hours before turning off the furnace. The temperature gradient in the dual-zone furnace was kept at 5 °C/cm.

For $Sr_{8/7}TiS_3$, single crystal growth was done in two steps. First, polycrystalline powders of $Sr_{8/7}TiS_3$ were synthesized in a $CS_2$ annealing setup. Strontium titanate powders (Thermo Fisher, 99%) were loosely packed in an alumina furnace and loaded into a tube furnace. The furnace was heated to 800 °C, using a ramp rate of 10 °C/min, and held for 24 hours. During the reaction, argon was bubbled through $CS_2$ (Alfa Aesar, 99.9 %) with a flow rate of 12 sccm. After the hold time, the furnace was turned off, and the sample was allowed to cool down naturally. For single crystal growth using KI as flux, 10 g pre-dried KI powder (Alfa Aesar, 99.9%) and 366 mg of $Sr_{8/7}TiS_3$ powder in a 1: 40 powder: flux ratio were added to a quartz ampoule. The quartz ampoule was evacuated and sealed using a blow torch and loaded into a 6" tube furnace in a vertical position. The ampoule was heated to 1050 °C in 20 hours and held at that temperature for 20 hours, followed by a slow down to 650 °C using a ramp rate of 1 °C/hour. After that, the furnace was turned off.

**DFT calculations**

Density-functional theory (DFT) calculations were performed using projector augmented-wave potentials[44] as implemented in the Vienna Ab initio Simulation Package (VASP)[45, 46]. The Perdew-Burke-Ernzerhof (PBE) functional within the generalized gradient approximation (GGA)[47] was used to describe the exchange-correlation interactions. A plane-wave basis set with an energy cutoff of 600 eV and $10^{-8}$ eV for the electronic convergence were applied. A $k$-point spacing of 0.025 Å$^{-1}$ was chosen for both the structure optimization and total-energy calculations. The crystal structures were optimized until all forces on the atoms were less than $10^{-4}$ eV/Å. To better describe the localization of Ti-$d$ electrons, we used the DFT + $U$ approach[48]. An effective on-site Hubbard $U$ = 3.0 eV was used for the Ti-$d$ electrons based on a previous study[14]. Furthermore, we considered magnetic configurations for all the modulated $Sr_xTiS_3$ structures. The special quasi-random structure (SQS) model implemented in the alloy theoretic automatic toolkit[49, 50] was used to generate best approximations of randomness in the paramagnetic configuration. The visualization of band-decomposed charge density was performed for valence electrons within 0.5 eV below the Fermi energy. The frequency-dependent dielectric function was calculated by with the LOPTICS tag in the independent particle approximation (IPA)[33] as implemented within VASP. Convergence tests for the number of empty conduction band states were performed based on a previous study[14]. Considering the calculation efficiency and accuracy, we set the total number of energy bands to be 2.5 times as many as the number of valence bands for the dielectric function calculations.

**STEM characterization**

We prepared [100]-oriented TEM specimens from two $Sr_xTiS_3$ crystals using Ar-ion milling. The specimen thinning process involved utilizing a 4 keV ion beam with an angle of incidence at 5°, followed by a 1 keV ion beam with an angle of incidence at 2°. Scanning transmission electron microscope (STEM) imaging was performed using an aberration-corrected Nion UltraSTEM 100 operated at 100 kV with a convergence semi-angle of 30 mrad. HAADF images were acquired using an annular dark-field detector with inner and outer collection semi-angles of 80 and 200 mrad, respectively. To improve the signal-to-noise ratio and correct the image drift, we recorded a sequence of 20 HAADF images with fast scan (dwell time of 1 μs), to enable post-processing correction of sample drift and image registration.

Energy-dispersive X-ray spectroscopy (EDS) imaging in STEM mode was performed using a double-silicon drift detector (JED-2300T, Jeol Ltd.) equipped in a STEM instrument (ARM200CF, Jeol Ltd.). Each EDS detector has an effective X-ray sensing area of 100 mm$^2$ and provides a large effective solid angle of ~1.2 sr, which guarantees ~10% collection efficiency for the generated X-ray signals (4π sr).

To interpret the intensity variation in a STEM image, multi-slice simulations were carried out on the structures of $Sr_xTiS_3$. The structures of $Sr_xTiS_3$ were obtained after structural optimization using DFT. STEM-HAADF simulations were performed using the multi-slice method as implemented in μSTEM[51]. The sample thickness was set to 15 nm and the defocus value was set to 10 Å to obtain good agreement in intensity profiles with the experimental data. We performed the simulations using an aberration-free probe with an accelerating voltage of 100 kV and a convergence semi-angle of 30 mrad. The inner and outer collection angles for the HAADF detector were set to 80 and 200 mrad, respectively.

**Fourier Transform Infrared Spectroscopy (FTIR)**

Infrared spectroscopy was performed using a Fourier-transform infrared spectrometer (Bruker Vertex 70) connected to an infrared microscope (Hyperion 2000), using a Globar source, a potassium bromide beam splitter, and a mercury-cadmium-telluride (MCT) detector. The single crystal sample was suspended in air for the measurement. A ZnSe holographic wire grid polarizer was used for controlling the polarization of the incident light. Normal incidence reflection measurements were taken using a 15× Cassegrain microscope objective (numerical aperture = 0.4). A gold mirror was used for background measurement. Reflection spectra modelling details can be found in *Supporting Information, Section VII*.

# ACKNOWLEDGEMENTS

This work was supported by the National Science Foundation (NSF) through grants DMR-2122070 (G.Y.J., R.M.), 2122071, and 2145797 (G.R., R.M.). The crystal growth and processing capabilities used were supported in part by ONR grant with award number N00014-23-1-2818, and the polarization-resolved spectroscopy studies and tools were supported in part by Army Research Office grants with grant nos. W911NF-24-1-0164 and W911NF-25-1-0022, respectively. The Microscopy work was conducted as part of a user project at the Center for Nanophase Materials Sciences (CNMS), which is a DOE Office of Science User Facility using instrumentation within ORNL's Materials Characterization Core provided by UT-Battelle, LLC, under Contract No. DE-AC05-00OR22725 with the DOE and sponsored by the Laboratory Directed Research and Development Program of Oak Ridge National Laboratory, managed by UT-Battelle, LLC, for the U.S. Department of Energy. This work used computational resources through allocation DMR160007 from the Advanced Cyberinfrastructure Coordination Ecosystem: Services & Support (ACCESS) program, which is supported by NSF grants # 2138259, #2138286, #2138307, #2137603, and #2138296. Y.-M.K. acknowledges the support of the National Research Foundation of Korea (NRF) grant (RS-2023-NR076943) funded by the Korean government in Korea.

## Author contributions

G.R. and R.M. conceived the idea and designed the experiments. G.R. performed the theoretical calculations with assistance from G.Y.J. and R.M. S.S., H.C., B.Z., K.Y. and J.R. synthesized the single crystals of $Sr_xTiS_3$. G.R. carried out STEM experiments and their analyses under the supervision of A.R.L., J.A.H., M.C., and R.M. W.C. and Y.M.K. performed EDS-STEM measurements. S.S. and J.R. performed FTIR measurements. G.R. and R.M. drafted the manuscript with edits from all the coauthors.

## Competing interests

The authors declare no conflict of interest.

# Supporting Information

# Towards Atomic-Scale Control over Structural Modulations in Quasi-1D Chalcogenides for Colossal Optical Anisotropy

Guodong Ren[1*#], Shantanu Singh[2,3#], Gwan Yeong Jung[4#], Wooseon Choi[5], Boyang Zhao[2], Kevin Ye[2], Huandong Chen[2], Andrew R. Lupini[6], Miaofang Chi[6], Jordan A. Hachtel[6], Young-Min Kim[5,7], Jayakanth Ravichandran[2,3,8*], Rohan Mishra[1,4*]

[1] *Institute of Materials Science and Engineering, Washington University in St. Louis, St. Louis, MO 63130, USA*

[2] *Mork Family Department of Chemical Engineering and Materials Science, University of Southern California, Los Angeles, CA 90089, USA*

[3] *Core Center of Excellence in Nano Imaging, University of Southern California, Los Angeles, CA 90089, USA*

[4] *Department of Mechanical Engineering and Material Science, Washington University in St. Louis, St. Louis, MO 63130, USA*

[5] *Department of Energy Science, Sungkyunkwan University, Suwon 16419, Republic of Korea*

[6] *Center for Nanophase Materials Sciences, Oak Ridge National Laboratory, Oak Ridge, TN 37830, USA*

[7] *Center for 2D Quantum Heterostructures, Institute for Basic Science, Suwon 16419, Republic of Korea*

[8] *Ming Hsieh Department of Electrical Engineering, University of Southern California, Los Angeles, CA 90089, USA*

[#]These authors contributed equally: Guodong Ren, Shantanu Singh, Gwan Yeong Jung

*Corresponding authors: Guodong Ren (guodong.ren@wustl.edu); Jayakanth Ravichandran (j.ravichandran@usc.edu); Rohan Mishra (rmishra@wustl.edu)

# Contents

## Section I. Structural modulation in Sr$_x$TiS$_3$ compounds

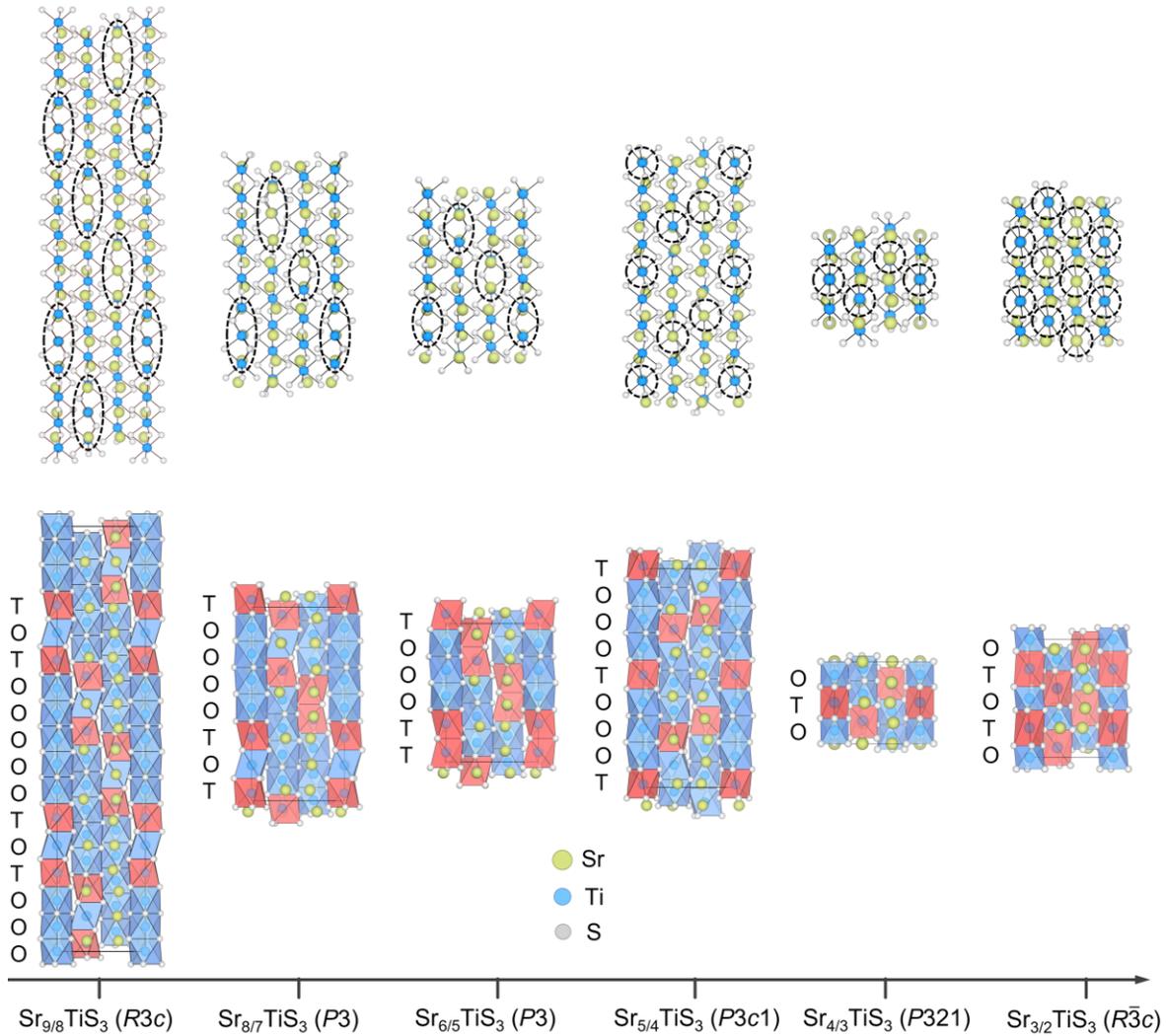

**Figure S1. Schematics of Sr$_x$TiS$_3$ structures with 1≤$x$≤1.5 viewed along [100] axis.** Top panel: Schematic representing modulated Sr$_x$TiS$_3$. The dashed ellipses indicate the periodic local blocks, where Ti atoms overlap with Sr atoms, when projected along [100] zone axis. These overlapping atomic columns display a higher intensity compared with non-overlapping atomic columns in HAADF-SETM images. Bottom panel: Modulated structures of Sr$_x$TiS$_3$ compounds viewed from [100] zone axis. The octahedral (O) and distorted trigonal prismatic (T) TiS$_6$ units are highlighted in blue and red, respectively.

Along *c*-axis, the commensurate modulated Sr$_x$TiS$_3$ only show one type of overlapping blocks: single block (Sr$_{5/4}$TiS$_3$, Sr$_{4/3}$TiS$_3$, Sr$_{3/2}$TiS$_3$), double block (Sr$_{8/7}$TiS$_3$, Sr$_{6/5}$TiS$_3$) and triple block (Sr$_{9/8}$TiS$_3$, Sr$_{8/7}$TiS$_3$). Uniquely among all the commensurate modulated structures, Sr$_{8/7}$TiS$_3$ lattice consists of two distinct stacking sequences among two TiS$_6$ chains: [−(T−O−T)−(O)$_4$−] and [−(O)$_5$−(T)$_2$−], which shows overlapping triplets and doublets within the unit cell.

**Section II. Thermodynamic stability through convex-hull analysis**

We evaluated the thermodynamic stability of Sr$_x$TiS$_3$ structures with different chemical compositions by constructing the convex hull with respect to the formation energy of their possible decomposition products. The convex hull connects phases that have a formation energy lower than any other phase or linear combination of phases at the respective compositions. For the Sr-Ti-S system, we calculated the total energy ($E_{total}$) of all the thermodynamically stable phases ($E_{hull}=0$) available in the Materials Project[1] as listed in the Table. S1. We constructed the convex hull through the grand canonical linear programming (GCLP) method[2]. The GCLP minimizes the free energy of a mixture at a given composition in the Sr-Ti-S phase space to identify the combination of thermodynamically equilibrium phases:

$$\Delta G = \sum_i f_i \Delta E_f^i,$$

where $\Delta G$ is the free energy of the compound with the desired composition, $f_i$ is the molar fraction of competing phases, $\Delta E_f^i$ is the formation energy of competing phases. We calculated $\Delta E_f^i$ of each phase using:

$$\Delta E_f^i = E_{Sr_xTi_yS_z} - xE_{Sr} - yE_{Ti} - zE_S,$$

where $E_{Sr_xTi_yS_z}$, $E_{Sr}$, $E_{Ti}$ and $E_S$ are the total energy of each phase from DFT calculations; while $x$, $y$ and $z$ are atomic fractions of each element in the compounds.

For Sr$_x$TiS$_3$ compounds discussed in this work, we find the corresponding decomposition products that minimize the total energy as written below:

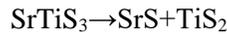
$$SrTiS_3 \rightarrow SrS + TiS_2$$

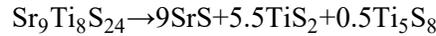
$$Sr_9Ti_8S_{24} \rightarrow 9SrS + 5.5TiS_2 + 0.5Ti_5S_8$$

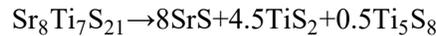
$$Sr_8Ti_7S_{21} \rightarrow 8SrS + 4.5TiS_2 + 0.5Ti_5S_8$$

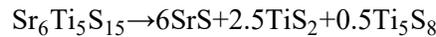
$$Sr_6Ti_5S_{15} \rightarrow 6SrS + 2.5TiS_2 + 0.5Ti_5S_8$$

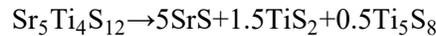
$$Sr_5Ti_4S_{12} \rightarrow 5SrS + 1.5TiS_2 + 0.5Ti_5S_8$$

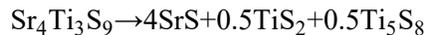
$$Sr_4Ti_3S_9 \rightarrow 4SrS + 0.5TiS_2 + 0.5Ti_5S_8$$

$$Sr_3Ti_2S_6 \rightarrow 3SrS + 0.333TiS + 0.333Ti_5S_8$$

As shown in the tabulated dataset Table. S1 for the Sr-Ti-S system, modulated $Sr_xTiS_3$ structures with $1<x<1.5$, are thermodynamically stable with formation energy on the hull ($E_{hull}=0$), while $Sr_{3/2}TiS_3$ ($x$=1.5) and stoichiometric $SrTiS_3$ ($x$=1) are metastable having formation energy above hull.

**Supplementary Table. S1** Database for convex hull construction

| Formula | Space group | $E_{total}$/eV | Atoms/unit | $\Delta E_f$ (eV/atom) | $E_{hull}$ (meV/atom) |
|---|---|---|---|---|---|
| Sr | $R\bar{3}m$ | -5.017 | 3 | 0 | |
| Ti | $P6/mmm$ | -16.578 | 3 | 0 | |
| S | $P2/c$ | -132.103 | 32 | 0 | |
| SrS | $Fm\bar{3}m$ | -10.113 | 2 | -2.156 | |
| $SrS_3$ | $Aba2$ | -36.283 | 8 | -1.021 | |
| $Ti_2S$ | $Pnnm$ | -218.902 | 36 | -1.020 | |
| $Ti_2S_3$ | $C2/m$ | -61.002 | 10 | -1.413 | |
| $Ti_5S_8$ | $C2/m$ | -79.046 | 13 | -1.415 | |
| $Ti_7S_{12}$ | $P\bar{1}$ | -229.359 | 38 | -1.393 | |
| TiS | $P\bar{6}m2$ | -12.506 | 2 | -1.426 | |
| $TiS_2$ | $P\bar{3}m1$ | -17.849 | 3 | -1.355 | |
| $TiS_3$ | $P2_1/m$ | -43.744 | 8 | -0.990 | |
| $SrTiS_3$ | $P2_1$ | -55.474 | 10 | -1.631 | 44.975 |
| $Sr_{9/8}TiS_3$ | $R3c$ | -457.439 | 82 | -1.717 | 0 |
| $Sr_{8/7}TiS_3$ | $P3$ | -602.518 | 108 | -1.725 | 0 |
| $Sr_{6/5}TiS_3$ | $P3$ | -435.035 | 78 | -1.747 | 0 |
| $Sr_{5/4}TiS_3$ | $P3c1$ | -702.425 | 126 | -1.765 | 0 |
| $Sr_{4/3}TiS_3$ | $P321$ | -266.998 | 48 | -1.786 | 0 |
| $Sr_{3/2}TiS_3$ | $R\bar{3}c$ | -364.326 | 66 | -1.807 | 12.498 |

## Section III. Theoretical calculation of complex dielectric function

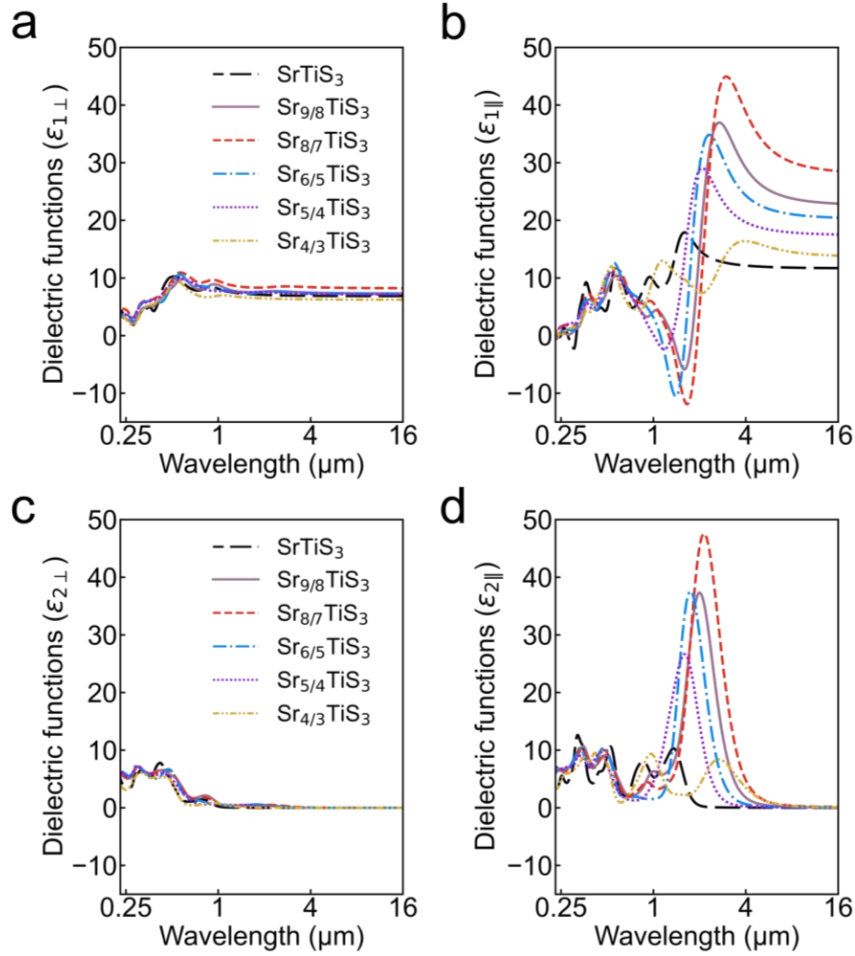

**Figure S2. Frequency-dependent dielectric functions of Sr$_x$TiS$_3$ calculated from DFT. a-b,** Real part of dielectric functions for polarization perpendicular ($\varepsilon_{1\perp}$) and parallel ($\varepsilon_{1\parallel}$) to the *c*-axis of Sr$_x$TiS$_3$ structures. **c-d,** Imaginary part of dielectric functions for polarization perpendicular ($\varepsilon_{2\perp}$) and parallel ($\varepsilon_{2\parallel}$) to the *c*-axis of Sr$_x$TiS$_3$ structures.

## Section IV. Theoretical calculation of electronic structures

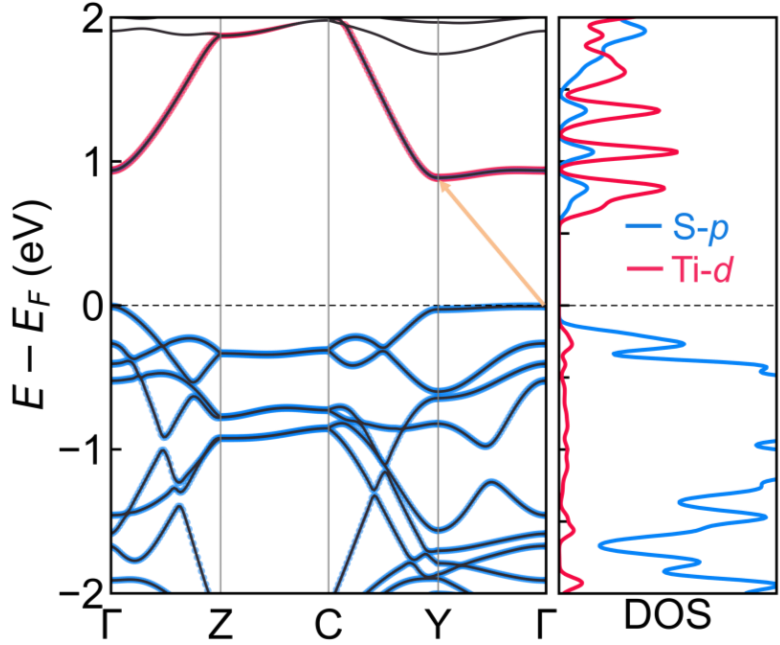

**Figure S3. Electronic structures of SrTiS$_3$ structures.** (Left panel) Band structure with projected wavefunction character and (right panel) element- and orbital-projected density of states (PDOS) showing the hybridization of S-$p$ and Ti-$d$ states in SrTiS$_3$. The orange arrows, connecting the valence band maximum (VBM) to the conduction band minimum (CBM), indicate the indirect bandgap in SrTiS$_3$.

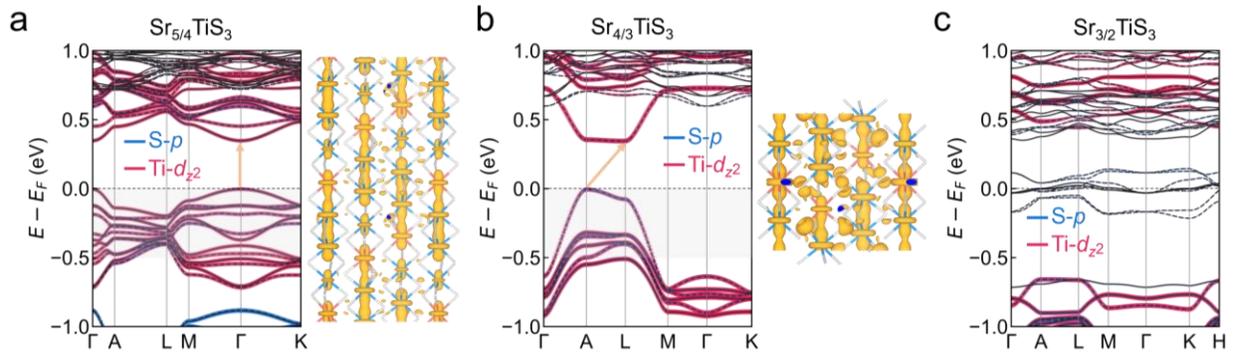

**Figure S4. Electronic structures of Sr$_x$TiS$_3$ ($x = [5/4, 4/3, 3/2]$) structures calculated from DFT+$U$ ($U$=3.0 eV). a-c,** Left panel: orbital-projected band structures for modulated Sr$_{5/4}$TiS$_3$ in **a**, Sr$_{4/3}$TiS$_3$ in **b**, and Sr$_{3/2}$TiS$_3$ in **c**. The thicker lines highlighted in red correspond to the contribution from Ti-$d_{z^2}$ states, while thicker lines in blue correspond to S-$p$ states. The Fermi energy is set to 0 eV. Right panel: spatial distribution of the valence electrons within 0.5 eV below the Fermi energy (shaded in gray in **a-c**). The isosurface is set to an electron density of 0.004 e/Å$^3$.

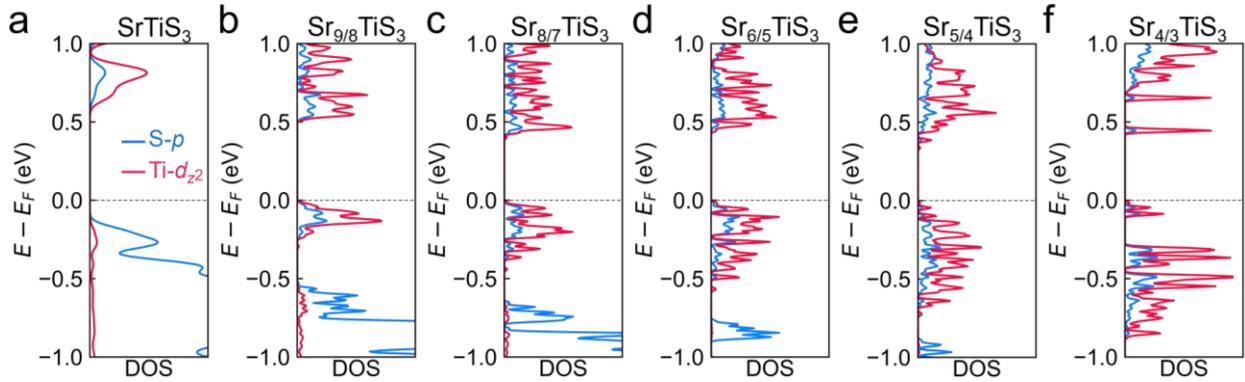

**Figure S5. Projected density of states of Sr$_x$TiS$_3$ structures.** As more Sr introduced into Sr$_x$TiS$_3$ structures, more Ti-$d_{z^2}$ states hybridized with S-$p$ orbital states become occupied by additional valence electrons from excess Sr atoms.

**Section V. Atomic-resolution STEM observation**

In some local regions of synthesized Sr$_{8/7}$TiS$_3$ crystals, we observe several brighter horizontal columns (perpendicular to the $c$-axis) that segment the crystal into different regions, as shown in the large field-of-view HAADF image in Figure S6a. From the corresponding FFT pattern, shown in the inset of Figure S6a, we observe a streaking of spots along the $c$-axis, which signifies local variation of modulation periodicity. To measure any variations in the chemical composition across these brighter horizontal atomic columns, we acquired atomic-resolution STEM-EDS images (Figure S6b-e). However, our STEM-EDS analysis did not reveal any discernible elemental aggregation across these brighter horizontal columns. More strikingly, we find a mixed arrangement of brighter doublets and triplets along the $c$-axis spaced by the brighter horizontal columns, as shown in the high-magnification HAADF image in Figure S7. The mixed arrangement of brighter doublets and triplets does not correspond to any commensurate modulations of the Sr$_x$TiS$_3$ lattice, based on our atomic-structure calculations, as illustrated in Figure 1 and Figure S1. The presence of brighter doublets suggests a local composition of either Sr$_{6/5}$TiS$_3$ or Sr$_{8/7}$TiS$_3$, as only these two structures have doublet blocks of –(T–O)–, as shown in Figure S1. Therefore, based on our compositional and structural analyses, we hypothesize that the brighter horizontal columns in these crystals could represent interfaces between different modulated structures.

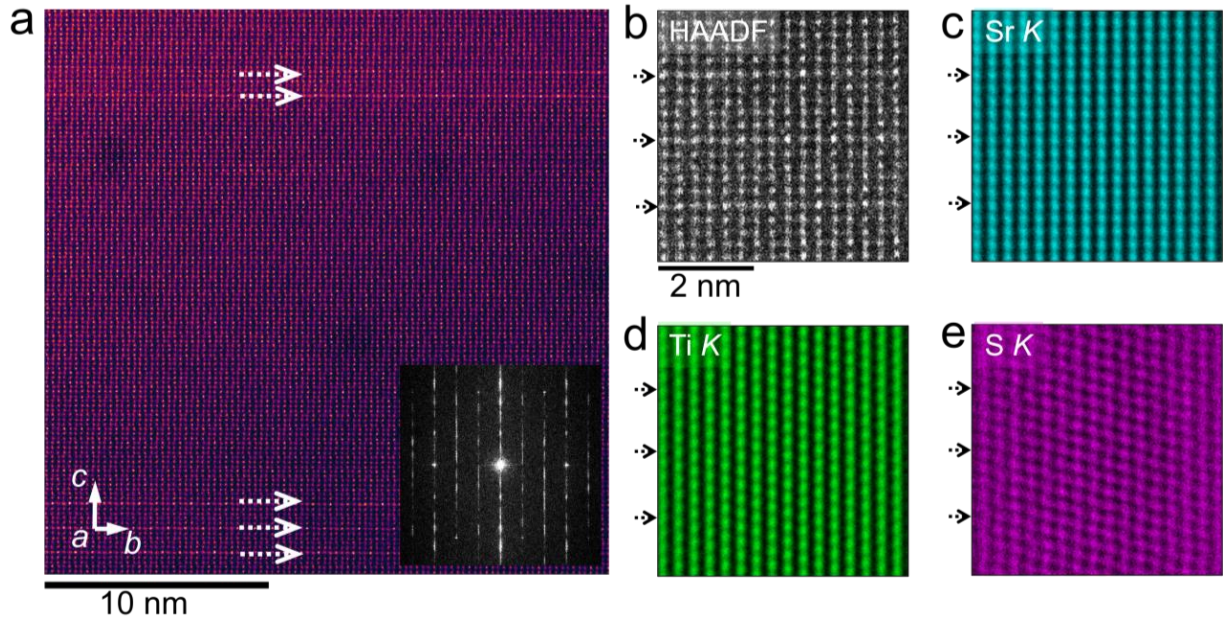

**Figure S6. Atomic-resolution STEM-EDS image of a synthesized $Sr_{8/7}TiS_3$ crystal along [100] axis. a,** Large field of view HAADF-STEM image showing horizontal coherent interfaces with brighter intensity marked with white arrows. The inset shows the corresponding FFT pattern with a clear streaking of the spots along the *c*-axis due to the variation of local modulation periodicity. **b,** Simultaneously acquired HAADF-STEM image showing the brighter lines perpendicular to *c*-axis. The black arrows at bottom indicate the location of brighter lines. **c-e,** Elemental mappings for Sr-*K* in **c**, Ti-*K* in **d** and S-*K* in **e**. There is no discernible elemental aggregation across these brighter horizontal lines shown in **a** and **b**.

We have investigated the occurrence of the local structural modulations in the flux-grown $Sr_{8/7}TiS_3$ crystals by performing a template-matching using a fast normalized cross-correlation method[3, 4]. Simulated HAADF-STEM images of $Sr_{6/5}TiS_3$ and $Sr_{8/7}TiS_3$ lattices, shown in Figure S7b, were used as the image templates to detect the local structural correlation across the HAADF-STEM image in Figure S7a. The spatial variation of structural similarities evaluated from template-matching across Figure S7a, is shown as color maps in Figure S7c for $Sr_{8/7}TiS_3$, and in Figure S7d for $Sr_{6/5}TiS_3$, respectively. The color bars in Figure S7(c-d) quantify the cross-correlation for $Sr_{8/7}TiS_3$ and $Sr_{6/5}TiS_3$ in the local regions, where the maximum value 1 shows a perfect match and minimum value 0 signifies no similarity. In Figure S7c, $Sr_{8/7}TiS_3$ lattice exhibits the highest probability to exist in the middle of Figure 5a between the brighter horizontal columns. While, in Figure S7d, $Sr_{6/5}TiS_3$ lattice is highly possible to exist in both the top and bottom regions of Figure S7a, which is complementary to the spatial distribution of $Sr_{8/7}TiS_3$ in Figure S7c.

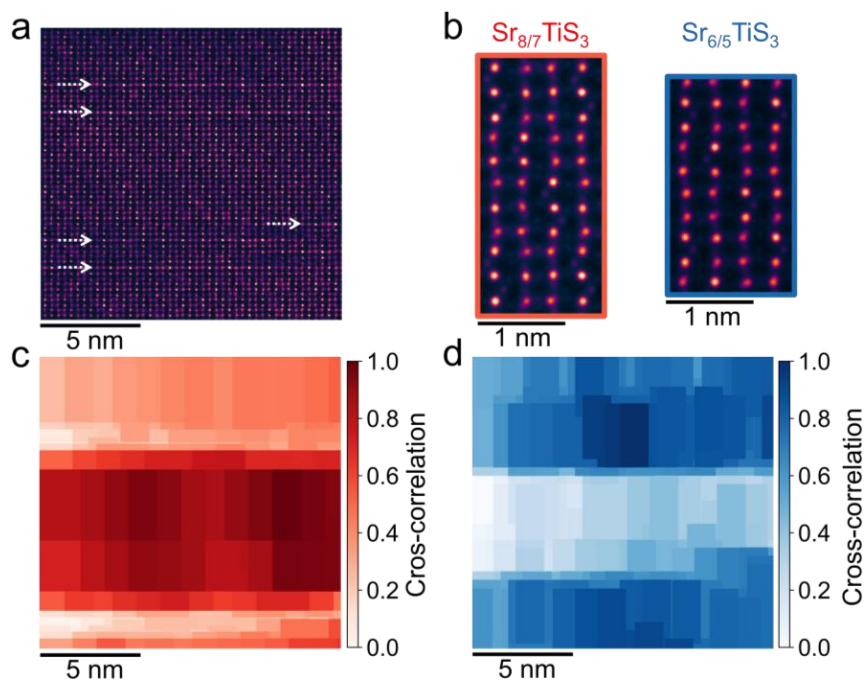

**Figure S7. Mixed modulation periodicity within an aperiodic Sr$_x$TiS$_3$ crystal synthesized with excess Sr in the precursor. a,** Atomic-resolution HAADF image of the Sr$_x$TiS$_3$ crystal viewed along the [100] zone axis showing brighter horizontal atomic columns marked with white arrows. **b,** Simulated HAADF images of Sr$_{8/7}$TiS$_3$ (left panel) and Sr$_{6/5}$TiS$_3$ (right panel), viewed along the [100] axis. **c-d**, Template-matching using fast normalized cross-correlation shows regions with Sr$_{8/7}$TiS$_3$ in **c** and Sr$_{6/5}$TiS$_3$ in **d**.

We further compared the intensity and the spacing between atomic columns in the experimental and the simulated HAADF images. In Figure S8, we observe an excellent agreement between experimental result and simulated STEM image of Sr$_{6/5}$TiS$_3$ lattice by comparing the intensity variation across the atomic columns.

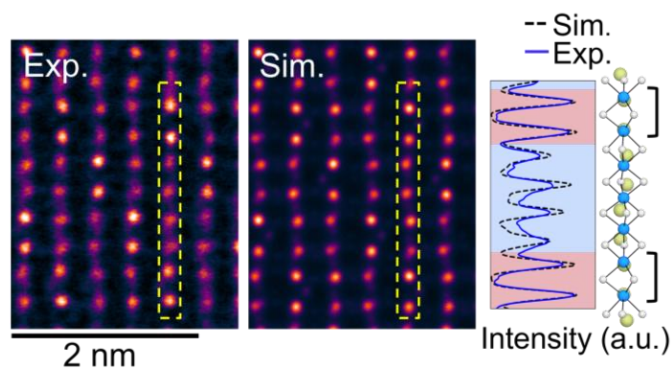

**Figure S8. HAADF-STEM images viewed along [100] axis showing the mixed modulation periodicities.** Modulated structure extracted from local region where Sr$_{6/5}$TiS$_3$ lattice shows the highest similarity evaluated from template-matching analysis. Line profile across the experimental and simulated

STEM images compares the intensity variation across two neighboring atomic columns as enclosed by dash boxes.

## Section VI. Fourier Transform Infrared Spectroscopy (FTIR)

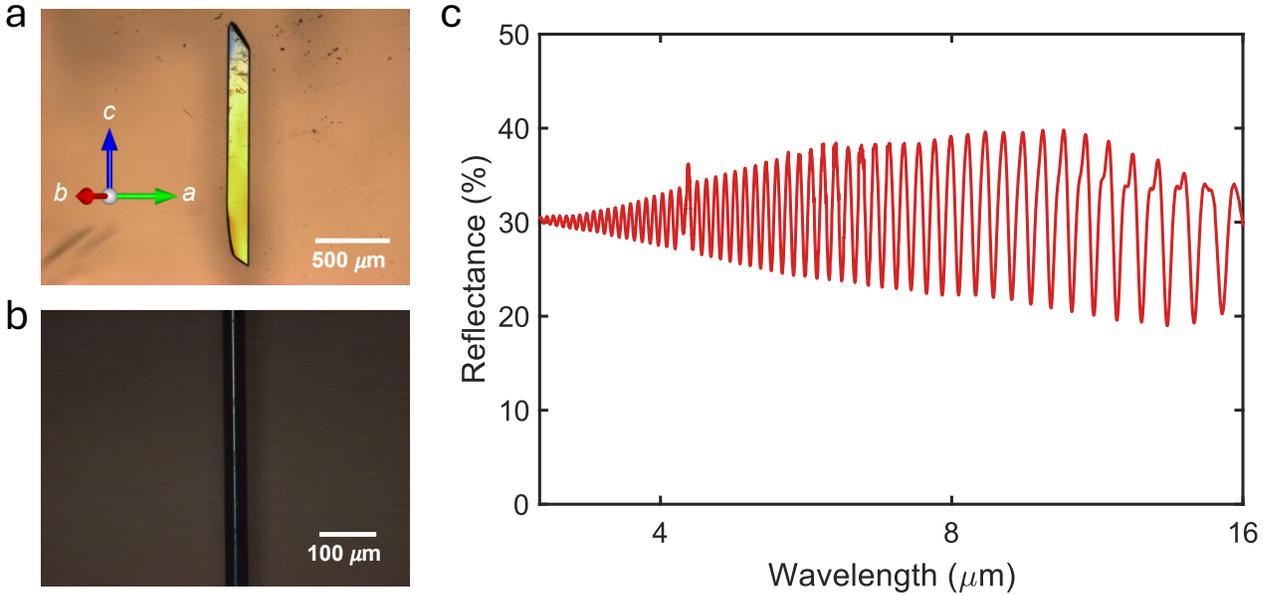

**Figure S9: Optical images and Unpolarized FTIR Reflectance spectrum. a,** Optical image of the $Sr_{8/7}TiS_3$ single crystal used for FTIR measurement with crystallographic orientation labelled. **b,** Cross section of the $Sr_{8/7}TiS_3$ crystal. **c,** Unpolarized reflectance spectra for $Sr_{8/7}TiS_3$, measured upto the long-wavelength infrared regime (the wavelength axis is plotted in log scale).

The crystal used for FTIR measurements was ~133 μm in lateral dimensions and ~40.7 μm thick, as shown in **Figure S9**. This thickness was used to determine the FFT-derived refractive index values. The unpolarized FTIR spectrum measured for $Sr_{8/7}TiS_3$ is shown in **Figure S9c**. The spectrum demonstrates two periodicities for the interference fringes, corresponding to the refractive indices parallel and perpendicular to the *c*-axis. For doing FFT, the wavelength ranges of 8 – 16 μm and 10 – 16 μm were chosen for approximating refractive indices perpendicular and parallel to the *c*-axis, respectively. These ranges were selected based on DFT calculations that indicate the material is sufficiently transparent (with k < 0.02 and α < 250 cm⁻¹), and the refractive index remains constant (see **Figure 2d**). Doing FFT over a wavelength range where the refractive index is relatively constant can give a very good estimate of the refractive index ratios. This method has been employed to accurately determine refractive index values for $BaTiSe_3$ single crystals.[5]

## Section VII. Refractive Index Modeling based on Sellmeier Equation

To validate the refractive index values derived from the FFT analysis of the polarization-resolved FTIR spectra, a two-term Sellmeier equation was used to estimate the wavelength-dispersion of the refractive index in the transparent region (the same wavelength range used for the FFT analysis) for both ordinary and extraordinary axis. The following equation was used to model the wavelength-dependent refractive index:

$$n^2(\lambda) = A + \frac{B\lambda^2}{\lambda^2 - C}$$

Here, $n$ is the real part of the refractive index, $\lambda$ is the wavelength in $\mu$m, and $A$, $B$, and $C$ are the fitted Sellmeier coefficients. This form of the Sellmeier equation was chosen to minimize the number of free parameters while adequately fitting the fringe periodicity. The full reflection spectrum was modeled using the transfer matrix method (TMM), considering all reflections for a flat crystal suspended in air. The wavelength-dispersion of the refractive index from the Sellmeier equation was taken into account in the TMM to model the reflectivity fringes. In **Figures S10a and b**, the modeled reflectance spectrum (for both perpendicular and parallel to the *c*-axis) is compared to the experimental data. The simulated spectra were normalized to the data to compare their periodicities. **Figures S10c and d** show the corresponding refractive index dispersion for both directions as modeled from the Sellmeier equation. The determined Sellmeier constants for the dispersion perpendicular to the *c*-axis are: A = 6.825, B = 0.01, and C = 8 $\mu m^2$. For the dispersion parallel to the *c*-axis, the constants are: A = 22.5, B = 3.2, C = 8.7 $\mu m^2$. It should be noted that the modeled values agree quite well with the previously mentioned refractive indices derived from the FFT analysis, as shown in **Figure 5b** (~2.6 for perpendicular to *c*-axis and ~5.1 for parallel to *c*-axis). Our modeling also confirms that the approximation of a relatively constant refractive index in the transparent regime is valid in this scenario.

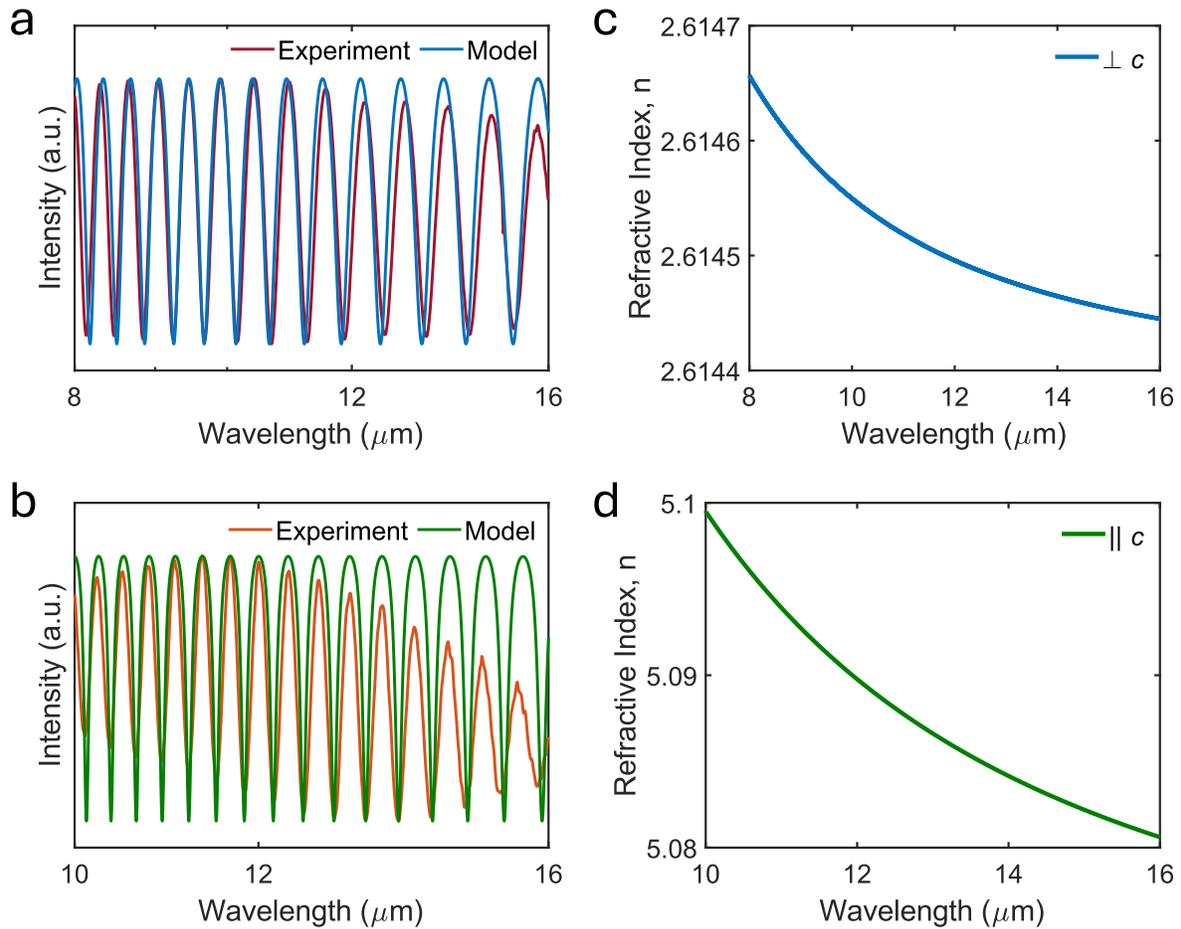

**Figure S10: Sellmeier equation-based modeling of Refractive index.** Comparison of modelled reflectance spectrum with the experimental reflectance spectrum for polarization **a**, perpendicular to *c*-axis, and **b,** parallel to *c*-axis. The spectra have been normalized in the displayed wavelength range. Refractive index dispersion derived from the model for **c**, perpendicular to *c*-axis, and **d,** parallel to *c*-axis.